\documentclass[twocolumn,pre,showpacs]{revtex4-1}
\usepackage{epsfig,amsmath,multirow}
\def\vec#1{\mathbf{#1}}
\def\vecc#1{{#1}}

\begin{document}

\title{Internal degrees of freedom and transport of benzene on graphite}
\author{Astrid S. de Wijn}
\affiliation{Radboud University Nijmegen, Institute for Molecules and Materials, Heyendaalseweg 135, 6525AJ Nijmegen, the Netherlands}

\begin{abstract}
In this paper, the chaotic internal degrees of freedom of a benzene molecule adsorbed on a graphite substrate, their interplay with thermal noise, and their effects on the diffusion and drift are investigated analytically by making use of the presence of two different time scales as well as by molecular-dynamics simulations.
The effects of thermal noise are investigated, and it is found that noise does not significantly alter the dynamics of the internal degrees of freedom, yet affects the friction and diffusion of the center of mass.
Qualitative and quantitative theoretical predictions for the friction and diffusion of the molecule on the substrate are made
and are compared to molecular-dynamics simulations.
Contributions to the friction and diffusion from the finite heat bath as well as the slow dynamics of the center of mass are formally identified.
It is shown that the torsion in benzene, which dominates the nonlinear coupling, significantly affects the friction of the molecule on the surface.
The results compare favorably with recent results from He/neutron spin echo experiments on this system.
Based on the analytical and numerical results, some suggestions are made for experimental conditions under which the effects of internal degrees of freedom might be observable.
\end{abstract}

\maketitle

\section{Introduction}

The presence of internal degrees of freedom (d.o.f.), especially rotation, has been connected experimentally and theoretically to the diffusion and friction of nanoscale objects on surfaces (see, for instance, Refs.~\cite{torqueandtwist,onsgraphiteflakes,rotar,makingmolecularmachineswork,onswdps13}).
Recent He/neutron spin echo experiments~\cite{holly,fouquetscatteringquestions,jardinereview} of benzene molecules adsorbed on graphite have shown that the friction is surprisingly high and it has been conjectured that this could be related to the internal d.o.f.\ of the molecule.

Several possible links can be identified between internal d.o.f.\ and transport.
The internal configuration of the molecules affects the interaction with the substrate, and therefore transport (see, for instance, Refs.~\cite{longjumps,mdcarboxylicacid}).
However, the deterministic internal d.o.f.\ can also act as a source of noise~\cite{onsbball,onswdps13}, if their dynamics are sufficiently fast and chaotic.
In this paper, I investigate the chaotic internal d.o.f.\ of a benzene molecule on a graphite substrate, their interplay with thermal noise, and their effects on the diffusion and friction, by making use of the presence of two different time scales.
The fast chaotic internal degrees of freedom act as a finite heat bath which couples to the slow motion of the center of mass.
I show that under realistic conditions, the time-scale separation theory of Ref.~\cite{tss1} is valid and links the dynamics of the internal degrees of freedom directly and quantitatively to the diffusion and friction of the molecule on the substrate. 
The theoretical results compare well, qualitatively and quantitatively, with simulations and experiments.
One of the goals of this research is to construct systems where the friction or diffusion can be controlled by external means, by manipulating the internal degrees of freedom.

In Sec.~\ref{sec:tss}, the concept of time-scale separation in the context of nonlinear dynamics and surface adsorbates is introduced briefly.
In Sec.~\ref{sec:model}, the atomistic model for benzene on graphite used in this work is described.
In Sec.~\ref{sec:effective}, the time-scale separation theory is applied to the benzene on graphite model with thermal noise and expressions for the effective interactions are derived.
It is also shown that thermal noise affects the dynamics of the internal d.o.f.\ weakly, but can still significantly affect the effective momentum diffusion and friction.
A relation between the effective momentum diffusion and friction is also derived.
Molecular-dynamics simulations of the system are described in Sec.~\ref{sec:md}, in which specific d.o.f.\ can be frozen to investigate their effects, and thermal noise can be introduced slowly.
The theoretical expressions are evaluated numerically and shown to compare favorably to simulation results and experimental results of benzene on graphite in Sec.~\ref{sec:comparison}.
Finally, in Sec.~\ref{sec:discussion}, conclusions are drawn, and suggestions are made for experimental conditions under which the effects of chaotic internal d.o.f.\ might be observable.

\section{\label{sec:tss}Time-scale separation}

We first consider the nature of the coupling between the internal d.o.f.\ and the center of mass.
In Refs.~\cite{onsbball,onswdps13}, the connections between transport and chaos in internal d.o.f.\ were outlined by taking advantage of the presence of multiple time scales.

When a molecule like benzene is adsorbed weakly (physisorbed) on a surface like graphite, it is not strongly distorted by the surface.
Accordingly, the internal forces are stronger than the forces exerted by the substrate and the time scales of the dynamics of the internal d.o.f.\ (for example vibrational modes) are much shorter than the time scales of the motion of the center of mass on the substrate.
Specifically, the equations of motion can be split into those for the slow motion of the center of mass $\vec{a}=(\vec{R},\vec{P})$ and those for the fast internal coordinates $\vec{b}=(\vec{q},\vec{p})$,
\begin{align}
\label{eq:tss1}
\dot{\vec{a}}& = \vec{f}(\vec{a},\vec{b})~,\\
\dot{\vec{b}}& = \frac1\epsilon \vec{g}(\vec{a},\vec{b})~,
\label{eq:tss2}
\end{align}
where $\vec{f}(\vec{a},\vec{b})$ and $\vec{g}(\vec{a},\vec{b})$ are both well-behaved functions of the same order of magnitude, and $\epsilon$ is the time-scale separation parameter ($0<\epsilon\ll 1$).

To simplify the description of the slow subsystem, one usually wishes to eliminate the fast variables.
Several strategies exist for doing this, depending on the dynamical properties of the fast degrees of freedom.
A (quasi)periodic fast subsystem can be averaged out~\cite{averaging}
while a highly disordered fast subsystem with many d.o.f.\ can be described as
an infinite heat bath \cite{vankampenadiabaticelim}.

The internal dynamics of complicated molecules, however, are neither many nor (quasi)periodic \cite{onswdps13}, but rather they are chaotic.
A fast chaotic subsystem acts as a finite heat bath, leading to stochastic driving and damping of the slow system \cite{bianucci,tss1,riegertnew,Beck}
(in this case the center of mass of the molecule on the substrate), with clear signatures of the finite size of the fast system, and the finite amount of energy stored in it.

When a dynamical system explores the entire phase space for almost all initial conditions, it is called ergodic.
When, in addition, sets of initial conditions are smeared out after a long time, it is referred to as mixing.  (See, for instance Refs.~\cite{Arnoldergodic,bobsboek}.)
If the fast subsystem has this property, its long-time behavior is independent of the initial conditions.
If correlation in the fast subsystem also decays exponentially, then the dynamics of the slow coordinates $\vec{a}=(\vec{R},\vec{P})$ can be described by a Fokker-Planck equation for their probability density $\rho(\vec{a},t)$ \cite{tss1},
\begin{eqnarray}
\label{eq:fokkerplanck}
\frac{\partial \rho (\vec{a},t)}{\partial t} =&
\displaystyle \sum_{\mu\nu} \frac{\partial}{\partial \vecc{a}_\mu}\, \frac{\partial}{\partial \vecc{a}_\nu} D^{(2)}_{\mu\nu}(\vec{a}) \rho(\vec{a},t)
\nonumber\\
&\null- \displaystyle \sum_\mu \frac{\partial}{\partial \vecc{a}_\mu} [D^{(1)}_\mu (\vec{a}) \rho(\vec{a},t)]
~,
\end{eqnarray}
where $D^{(1)}(\vec{a})$ is the $6$-dimensional effective drift vector and $D^{(2)}(\vec{a})$ is the $6\times 6$ effective momentum diffusion matrix.
The subscripts $\mu,\nu$ indicate the six components of the slow coordinates, 1, 2, 3 referring to the positions and 4, 5, 6 to the momenta.
The effective momentum diffusion and drift functions can be expressed in terms of averages over the invariant density of the fast subsystem for fixed slow coordinates, i.e. the probability density in the phase space of the fast subsystem which is left unchanged by the time evolution of the dynamics.
One finds~\cite{tss1}
\begin{align}
\label{eq:diffusiontss}
D^{(2)}_{\mu\nu}(\vec{a}) &= \int_0^\infty dt' \left\langle \delta \vecc{f}_\mu\left(\vec{a},\xi({t'},\vec{b};\vec{a})\right) \delta \vecc{f}_\nu(\vec{a},\vec{b})\right\rangle_\mathrm{E}\\
& = O(\epsilon)~,\displaybreak[0]\\
D^{(1)}_{\mu}(\vec{a}) &= \langle \vecc{f}_\mu (\vec{a},\vec{b})\rangle_\mathrm{E}\nonumber\\
\label{eq:drifttss}
&
\phantom{=}
\null + \sum_\nu\int_0^\infty dt' \left\langle \vecc{f}_\nu(\vec{a},\vec{b}) \frac{\partial \delta \vecc{f}_\mu(\vec{a},\xi({t'},\vec{b};\vec{a}))}{\partial \vecc{a}_\nu}\right\rangle_\mathrm{E}\displaybreak[0]
\\& = \langle \vecc{f}_\mu (\vec{a},\vec{b})\rangle_\mathrm{E} + O(\epsilon^2)~,
\end{align}
where $\langle . \rangle_\mathrm{E}$ is used to indicate an ensemble average over the invariant density,
 $\xi(t,\vec{b};\vec{a})$ is the solution of the equations of motion for the fast subsystem for fixed $\vec{a}$ and initial condition $\vec{b}$,
and
\begin{align}
\delta\vec{f}(\vec{a},\vec{b}) &= \vec{f}(\vec{a},\vec{b}) - \left\langle \vec{f}(\vec{a},\vec{b})\right\rangle_\mathrm{E}~.
\end{align}
The symbol $O$ is used here to indicate the order of magnitude of the next terms, which are not necessarily negligible.
It should be noted that typical low-dimensional Hamiltonian systems are not mixing
but have a complicated phase space, with regions of chaotic and quasi-periodic behavior~\cite{Arnoldergodic,ott,bobsboek,1dbball2}.
However, the mixing property is not strictly required for the expressions above~\cite{onsbball}.
Furthermore, in Sec.~\ref{sec:noise}, it is shown that the addition of the substrate temperature ensures both exponential decay of correlation in the fast subsystem and mixing.
 
In short, if the internal d.o.f.\ of a molecule are chaotic, then they act as a finite heat bath, providing noise to drive diffusion in the momentum [Eq.~(\ref{eq:diffusiontss})] and viscous friction [the second term in Eq.~(\ref{eq:drifttss})], affecting the diffusion and friction of the center of mass of the molecule on a surface~\cite{onsbball,onswdps13}.

\section{chaos in molecules: benzene on graphite\label{sec:model}}

This work is concerned with the time-scale separation in benzene molecules on graphite, a prototype system for molecular adsorption.
In order to investigate benzene, a model was developed in Ref.~\cite{onswdps13} which is particularly suitable for investigation of dynamical properties.

Though molecules are in principle quantum-mechanical, often they can be described classically by atomistic models.
Most typical atomistic force fields used for this (for example Tripos~5.2~\cite{tripos5.2}) contain predominantly harmonic terms.
Nevertheless, complicated geometries as well as torsion, both present in the internal degrees of freedom of benzene, lead to strong nonlinearities in the interaction and to chaos~\cite{onswdps13}.

\begin{figure}
\includegraphics[width=8.6cm]{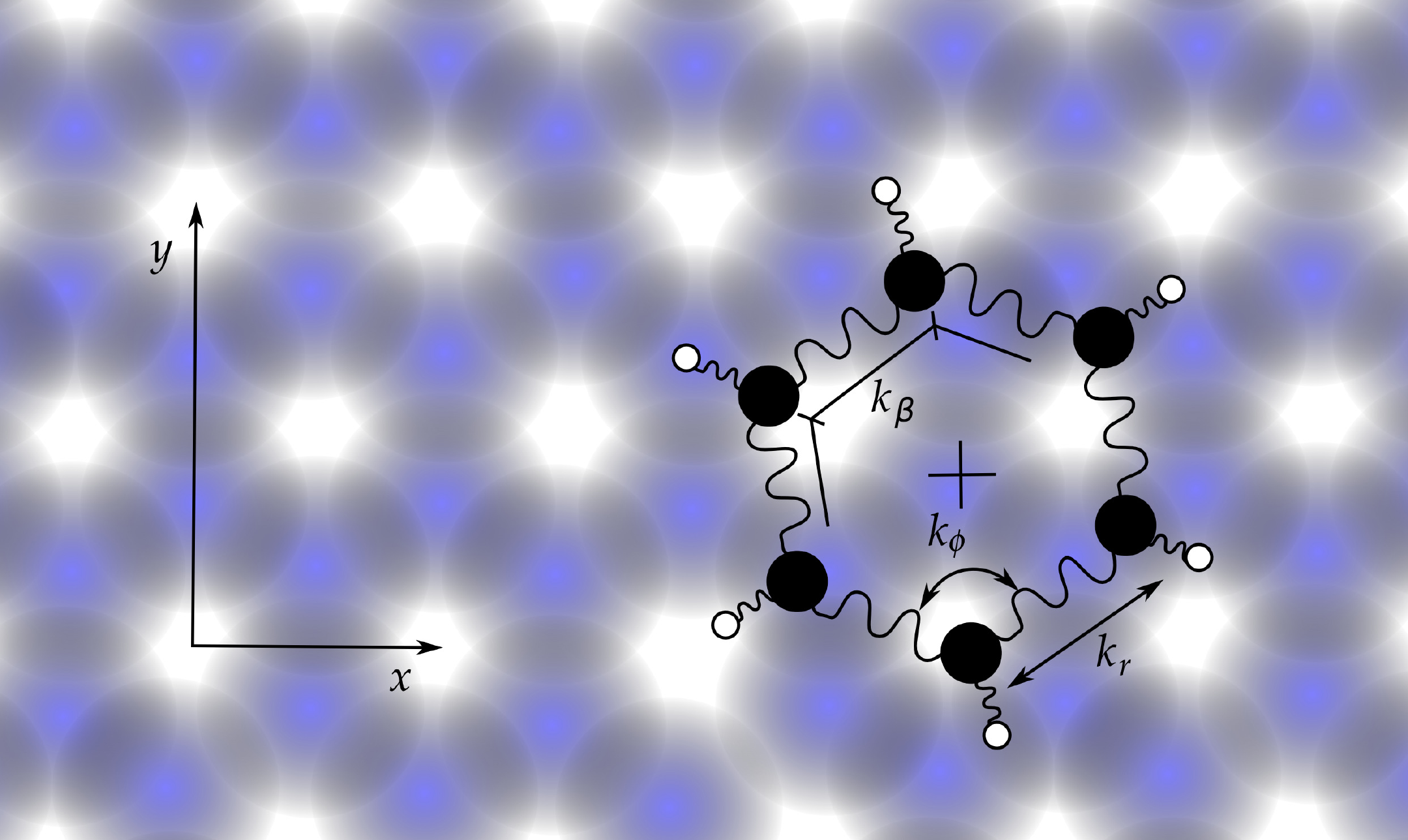}
\caption{
A schematic representation of a top view of a benzene molecule on a graphite substrate.
Because the C-H bonds must be described quantum mechanically, they must be eliminated from the equations of motion before a classical atomistic description can be used.
In the Tripos 5.2 force field three interaction parameters remain, the C-C stretching parameter $k_r$, the C-CH-C bending parameter $k_\phi$, and the C-CH-CH-C torsion parameter $k_\beta$.
\label{fig:cartoonbenzene}
}
\end{figure}

The total potential energy of a benzene molecule consists of contributions from bending, stretching, and torsion of various types of bonds.
However, due to the low mass of the hydrogen atoms, the energies associated with the vibrational modes involving the C-H bonds are high compared to the typical energy available at room temperature.
Consequently, the C-H bonds will always be in the ground state, and do not participate in the chaotic dynamics~\cite{onswdps13,ChaosBook}.
If an atomistic model is to be used to investigate the internal degrees of freedom of benzene, the C-H bonds must therefore be eliminated.
Here, we average them out using a mean-field approximation, to obtain a greatly simplified  Hamiltonian~\cite{onswdps13}.

Let $\vec{r}_i$ and $\tilde{\vec{p}}_i$ denote the position and momentum of the $i$-th CH complex, ordered in such a way that $i$ and $i+1$ are neighbors in the benzene ring.
Let $\phi_i$ be the angle between $\vec{r}_{i-1}-\vec{r}_i$ and $\vec{r}_{i+1}-\vec{r}_i$, and $\beta_i$ the torsion angle associated with the bonds between the $(i-1)$st, $i$th, $(i+1)$st, and \hbox{$(i+2)$nd} carbon atoms.
The potential energy of the benzene molecule can be written as
\begin{eqnarray}
V_\mathrm{benzene} (\vec{r}_1,\ldots,\vec{r}_6) &=&
{\textstyle \frac12} k_r \sum_{i=1}^6 (\|\vec{r}_{(i+1)(\mathrm{mod}~6)}-\vec{r}_{i}\|-r_0)^2
\nonumber\\
&&\null
+ {\textstyle \frac12} k_\phi \sum_{i=1}^6 (\phi_i-{\textstyle \frac23}\pi)^2
\nonumber\\
&&\null
+ k_\beta \sum_{i=1}^6
[1+\cos (2 \beta_i)]~,
\label{eq:Vbenzene}
\end{eqnarray}
where $k_r$ and $r_0$ are the C-C stretching force constant and equilibrium distance, while $k_\phi$ and $k_\beta$ are the effective bending force constant and the effective torsion constant.
The model system is shown in Fig.~\ref{fig:cartoonbenzene}.
In the mean-field approximation of Ref.~\cite{onswdps13} for the Tripos 5.2 force field~\cite{tripos5.2}, which will be used in this work, we have $r_0 = 1.47$~\AA,  $k_r = 60.7~\mathrm{eV}/$\AA$^2$, $k_\phi = 6.85~\mathrm{eV}/\mathrm{rad}^2$, and $k_\beta = 0.247~\mathrm{eV}$.
The typical energies (frequencies and thus time scales) can be estimated from the linearized system and
range between 0.048~eV and 0.24~eV, which is comparable to the fundamental vibrational frequencies of benzene.

Chaos can be quantified by the largest Lyapunov exponent, the rate at which infinitesimal perturbations expand.
If this is positive, then the system is chaotic.
The largest Lyapunov exponent of this model with energies equivalent to room temperature can be calculated from molecular-dynamics simulations and is $\lambda=0.5/$ps, regardless of the precise initial conditions~\cite{onswdps13}. 
The system is strongly chaotic and also (close to) ergodic and mixing.

\subsection{Graphite substrate with temperature}

Here, I consider benzene molecules adsorbed on a graphite substrate.
Graphite is frequently chosen as a substrate in experiments.
It is often modelled with two-dimensional sinusoidal potentials (see, for instance, Ref.~\cite{torqueandtwist}).
In this work a three-dimensional substrate potential is needed, and therefore the two-dimensional potential of a hexagonal graphite substrate in the $xy$ plane is extended by adding a harmonic term in the $z$ direction.
The coefficient for the harmonic term is chosen such that in a potential minimum the second derivative of the potential energy is the same in all three directions.
The potential to which the CH complex at position $\vec{r}=(x,y,z)$ is subject due to the substrate is given by~\cite{torqueandtwist}
\begin{eqnarray}
V_\mathrm{CH}(\vec{r}) = -\frac{2 V_\mathrm{c}}{9}\left[2\cos\left(\frac{2 \pi x}{a \sqrt{3}}\right)\cos\left(\frac{2 \pi y }{3 a}\right)
\right.\nonumber\\
\left.\null + \cos\left(\frac{4 \pi y }{3 a}\right)\right] + V_c \frac{8\pi^2}{27a^2} z^2~,
\label{eq:Vsubstrate}
\end{eqnarray}
where $V_\mathrm{c}= 25$~meV is the potential difference associated with the corrugation, and $a=1.42$~\AA\ is the in-layer inter-atomic distance of graphite.
The origin and direction of the $x$ and $y$ axes are indicated in Fig.~\ref{fig:cartoonbenzene}.
Note that the typical frequency of the dynamics of the center of mass in this potential (associated with the lowest energy mode at 0.38~meV) is more than two orders of magnitude lower than the lowest vibrational frequency of the molecule.

The Hamiltonian of the system can now be written as
\begin{align}
H = \frac1{2 m_\mathrm{CH}} \sum_{i=1}^6 \tilde{\vec{p}}_i^2 + V_\mathrm{benzene} (\vec{r}_1,\ldots,\vec{r}_6)\nonumber\\+ \sum_{i=1}^{6} V_\mathrm{CH}(\vec{r}_i)~,
\label{eq:hamiltonian}
\end{align}
where $\tilde{\vec{p}}_i$ is the momentum of the $i$-th CH complex, and $m_\mathrm{CH}$ is the mass of a CH complex.

In preliminary simulations without substrate thermal noise~\cite{onswdps13}, it was shown that in this system the chaos in the internal d.o.f.\ does indeed lead to diffusive behavior of the center of mass and that this diffusion is normal, as expected from the time-scale separation.

Because the substrate has a finite temperature, the molecule adsorbed on it is subjected to thermal fluctuations.
These are modelled here with Langevin dynamics applied to each CH complex: stochastic noise, driving the momentum with diffusion constant $D_T$, and viscous friction with momentum damping constant $\gamma_T$,
\begin{align}
\vec{F}_\mathrm{Langevin} = - \gamma_T m_\mathrm{CH} \dot{\vec{r}}_i + \boldsymbol{\eta}_i(t)~,
\label{eq:langevin}
\end{align}
where the amplitude of the noise must be chosen such that the expectation value of $\int_0^\infty dt\, \boldsymbol{\eta}_i(0) \cdot \boldsymbol{\eta}_i(t)$ is equal to $3 D_T$.
The friction and diffusion constants can be related to the temperature $T$ through
\begin{align}
\frac{D_T}{\gamma_T} = - m_\mathrm{CH} k_\mathrm{B} T~.
\label{eq:Dgamma}
\end{align}
The amplitude of the noise can be varied without changing the temperature, by increasing $\gamma_T$ proportionally with $D_\mathrm{T}$.
Typically, the friction coefficient $\gamma_T$ of a graphite substrate is expected to be around 1/ps.
However, in recent He/neutron spin echo experiments, it was found that the friction coefficient of benzene on graphite was unexpectedly high, $2.2\pm0.1$/ps~\cite{holly}.

\section{\label{sec:effective}Effective interaction}

The effective dynamics of the benzene molecule on the substrate are determined by the coupling and noise from the substrate and molecule.
In this section, the effective interaction is investigated analytically in more detail, and the various contributions are formally identified.

In order to apply the time-scale separation, and use the expressions of Eqs.~(\ref{eq:diffusiontss}) and~(\ref{eq:drifttss}) to determine the effects on the diffusion and friction of the molecule, the equations of motion must first be rewritten in the form of Eqs.~(\ref{eq:tss1}) and~(\ref{eq:tss2}),
by separating out the slow coordinates $\vec{R}, \vec{P}$ of the center of mass from the 15 fast internal d.o.f., which are denoted by $\vec{q}_1,\ldots, \vec{q}_5$, with associated momenta $\vec{p}_1,\ldots, \vec{p}_5$.
We have $\vec{P}=6m_\mathrm{CH}\dot{\vec{R}}$.

Let the lowest frequency of the linearized benzene molecule be denoted by $\omega_\mathrm{benzene}$, and the typical frequency of the substrate be denoted by $\omega_\mathrm{graphite}$.
The time-scale separation parameter is related to these two frequencies (given in the previous section) via
\begin{align}
\epsilon = \frac{\omega_\mathrm{graphite}}{\omega_\mathrm{benzene}} \approx 0.0078~.
\end{align}

Let the set of internal coordinates be denoted by
\begin{eqnarray}
\vec{q}_i=\frac{\vec{r}_i-\vec{R}}{\sqrt{\epsilon}}~,
\label{eq:qi}
\end{eqnarray}
for $i=1,\ldots,5$.
The corresponding momentum is
\begin{eqnarray}
\vec{p}_i= \sqrt{\epsilon}\, m_\mathrm{CH}(\dot{\vec{r}}_i-\dot{\vec{R}})~.
\end{eqnarray}
For convenience of notation, we shall use the notation $\vec{q}_6=-\sum_{i=1}^5 \vec{q}_i$ and $\vec{p}_6 = -\sum_{i=1}^5 \vec{p}_i$.
As the potential energy of benzene, Eq.~(\ref{eq:Vbenzene}), contains only differences of the positions of the CH complexes, and not $\vec{R}$ explicitly, Eq.~(\ref{eq:hamiltonian}) can be written as
\begin{align}
H  =& \frac{1}{12 m_\mathrm{CH}} \vec{P}^2 + \frac{1}{2 m_\mathrm{CH}\, \epsilon} \sum_{i=0}^6 {\dot{\vec{p}_i}}^2 \nonumber\\
 & \null + V_\mathrm{benzene} (\vec{q}_1/\sqrt{\epsilon},\ldots,\vec{q}_6/\sqrt{\epsilon})\nonumber\\
&\null+ \sum_{i=1}^{6} V_\mathrm{CH}(\vec{R}+\vec{q}_i/\sqrt{\epsilon})~,
\label{eq:benzenehamiltoniantss}
\end{align}

The equations of motion of the full system read,
\begin{align}
\label{eq:motionR}
\dot{\vec{R}} &= \frac{\vec{P}}{6 m_\mathrm{CH}}~,\displaybreak[0]\\
\label{eq:motionP}
\dot{\vec{P}} &= \null- \frac{\partial }{\partial \vec{R}} \sum_{i=1}^{6} V_\mathrm{CH}(\vec{R}+\vec{q}_i/\sqrt{\epsilon}) - \gamma_T {\vec{P}} +\sum_{i=0}^6 \boldsymbol{\eta}_i(t) ~,\displaybreak[0]\\
\label{eq:motionq}
\dot{\vec{q}}_i &= \frac{\vec{p}_i}{2 m_\mathrm{CH}\, \epsilon}~,\displaybreak[0]\\
\dot{\vec{p}}_i &=\null -\frac{\partial}{\partial\vec{q}_i}  V_\mathrm{benzene} (\vec{q}_1/\sqrt{\epsilon},\ldots,\vec{q}_6/\sqrt{\epsilon})
\nonumber\\ &\phantom{=}\null
- \frac{\partial }{\partial \vec{q}_i} V_\mathrm{CH}(\vec{R}+\vec{q}_i/\sqrt{\epsilon})
\nonumber\\ &\phantom{=}\null
- \gamma_T {\vec{p}}_i + \sqrt{\epsilon} \left(\boldsymbol{\eta}_i(t)-{\textstyle \frac16} \sum_{j=1}^6 \boldsymbol{\eta}_j(t)\right)~.
\label{eq:motionp}
\end{align}
From a comparison between the energy constants of the benzene molecule [Eq.~(\ref{eq:Vbenzene})], and the potential of the substrate [Eq.~(\ref{eq:Vsubstrate})], which is at least two orders of magnitude smaller, it can be concluded that the right-hand sides of Eqs.~(\ref{eq:motionq}) and~(\ref{eq:motionp}) are at least two orders of magnitude larger than the right-hand side of Eq.~(\ref{eq:motionR}) and the first term on the right-hand side of Eq.~(\ref{eq:motionP}).
Consequently, as long as the Langevin noise and friction due to the substrate temperature have the same slow time scales as the rest of the substrate, Eqs.~(\ref{eq:motionR}) and~(\ref{eq:motionP}) are of the form of Eq.~(\ref{eq:tss1}), while Eqs.~(\ref{eq:motionq}) and~(\ref{eq:motionp}) are of the form of Eq.~(\ref{eq:tss2}).

Equations~(\ref{eq:motionR}--\ref{eq:motionp}) can now be combined with Eqs.~(\ref{eq:diffusiontss}--\ref{eq:drifttss}) to calculate the effective interaction.
Let $\delta\vec{F}(\vec{R};\vec{q}_i)$ be the deviation of the total coupling $\vec{F}(\vec{R};\vec{q}_i)$ from the average,
\begin{align}
\delta\vec{F}(\vec{R};\vec{q}_i) = \vec{F}(\vec{R};\vec{q}_i) - \langle \vec{F}(\vec{R};\vec{q}_i) \rangle_\mathrm{E}~,\\
\vec{F}(\vec{R};\vec{q}_i) = - \frac{\partial }{\partial \vec{R}} \sum_{i=1}^{6} V_\mathrm{CH}(\vec{R}+\vec{q}_i/\sqrt{\epsilon})~.
\end{align}
Equation~(\ref{eq:motionR}) does not contain the fast variables, and therefore only the momentum of the center of mass is directly affected by the noise from the fast degrees of freedom.
This means that only components of $D^{(2)}_{\mu\nu}$ with $\mu,\nu>3$ can be nonzero.
The diffusion due to the noise from the two heat baths can then be written as
\begin{align}
\label{eq:diffusionPexpliciet}
D^{(2)}_{\mu\nu} & =
\left\{
\begin{array}{lll}
\tilde{D}_{\mu\nu}(\vec{R}) + \delta_{\mu\nu} D_T&& \mathrm{if}~\mu,\nu > 3\\
0&& \mathrm{otherwise}
\end{array}
\right.
~,\\
\tilde{D}_{\mu\nu}(\vec{R}) & = \int_0^\infty dt\, \langle \delta\vec{F}(\vec{R};\zeta(\vec{q}_i,t)) \delta\vec{F}(\vec{R};\vec{q}_i) \rangle_\mathrm{E}~,
\label{eq:difffinite}
\end{align}
where $\tilde{D}(\vec{R})$ is the effective momentum diffusion matrix due to the finite heat bath of the internal d.o.f.\ and $\delta_{\mu\nu}$ is the Kronecker delta.
The solution of equations of motion for the fast coordinates with initial conditions $\vec{q}_i$ and fixed slow coordinates is denoted by $\zeta(\vec{q}_i,t)$.

As the stochastic noise $\boldsymbol{\eta}_i(t)$ has zero average, the effective drift is
\begin{widetext}
\begin{align}
\label{eq:driftPexpliciet}
D^{(1)}_\mu & =
\displaystyle
\left\{\begin{array}{lll}
\displaystyle \frac{\vecc{P}_\mu}{6 m_\mathrm{CH}}&&\mathrm{if}~\mu \leq 3
\\[2ex]
\displaystyle \left\langle - \sum_{i=1}^6 \frac{\partial}{\partial \vecc{R}_\mu} V_\mathrm{CH}(\vec{R}+\vec{q}_i/\sqrt{\epsilon}) \right\rangle_\mathrm{E}
 - \vecc{P}_\mu\cdot \tilde{\gamma}(\vec{R}) - \gamma_T\vecc{P}_\mu
&& \mathrm{if}~\mu > 3 
\end{array}\right.~,
\end{align}
\end{widetext}
where $\tilde{\gamma}(\vec{R})$ is the $3\times3$ effective friction matrix, which is strictly positive definite.
It can be determined from the coupling and the decay of correlation in the internal d.o.f.\ by
\begin{align}
\tilde{\gamma}_{\mu\nu}(\vec{R}) = - \frac{1}{6 m_\mathrm{CH}}\int_0^\infty dt\, \left\langle \frac{ \delta\vec{F}_\mu(\vec{R};\zeta(\vec{q}_i,t)) }{\partial \vec{R}_\nu} \right\rangle_\mathrm{E}
\label{eq:fricfinite}
\end{align}

The first term in the second line on the right-hand side of Eq.~(\ref{eq:driftPexpliciet}), the average coupling due to the substrate potential, can be thought of as a configurational term.
It contains the invariant density of the internal d.o.f.\ through the ensemble average, which can be affected in detail by the dynamics of the internal degrees of freedom.
The second term is the viscous friction term due to the decay of correlation in the finite heat bath, while the third term is the viscous friction due to the coupling to the infinite heat bath of the substrate.
Both the friction and diffusion depend on the location on the substrate and the direction of the momentum.

Eqs.~(\ref{eq:difffinite}) -- (\ref{eq:fricfinite}) are a central result, giving theoretical expressions for the reduced dynamics of the center of mass.
In Sec.~\ref{sec:comparison}, these expressions will be evaluated and compared to simulations.

\subsection{The effects of thermal noise on the internal d.o.f.\label{sec:noise}}
The thermal noise from the substrate temperature not only affects the center of mass motion, but also the fast dynamics in the internal degrees of freedom.
As it is uncorrelated Gaussian white noise, it ensures exponential decay of correlation in the fast subsystem.

Because the coupling to the substrate is weak compared to the coupling between internal d.o.f., the fast subsystem is essentially subject to very weak Gaussian white noise and weak damping [see Eq.~(\ref{eq:motionp})].
Consequently, the dynamics remain similar to the dynamics without noise.

Nevertheless, the noise still has important qualitative and qualitative effects on the invariant density, and thus indirectly on the ensemble averages and the effective momentum diffusion and friction.
In particular, the noise smears out the dynamics and structures in the phase space~\cite{ott}.
This is important, because the results of Ref.~\cite{tss1} require that the fast subsystem be mixing, while low-dimensional Hamiltonian systems are, in general, not.
However, with a suitable realization of the Gaussian noise, any point in phase space can be reached in finite time from any initial condition.
In other words, the entire phase space is explored in finite time.
Together with the smooth dynamics of the benzene molecules, the above guarantees mixing.
Consequently, a unique invariant density exists, and the conditions of the time-scale separation theory are met.

In the presence of thermal noise, the fast subsystem is simply coupled weakly to an infinite heat bath, with a temperature equal to the substrate temperature $T$.
In the absence of other external driving, the ensemble average becomes equal to the thermal average, with the invariant density equal to the Gibbs distribution,
\begin{align}
\label{eq:thermalaverage}
&\langle A(\vec{R},\vec{P},\vec{q}_1,\ldots,\vec{p}_5) \rangle_\mathrm{E}\nonumber\\=
&\int d\vec{q}_1 \ldots d\vec{p}_5 \exp\left(-\frac{H}{k_\mathrm{B} T}\right) A(\vec{R},\vec{P},\vec{q}_1,\ldots,\vec{p}_5) ~.
\end{align}
Weak noise can change the invariant density significantly, and therefore the reduced slow dynamics through the average coupling, as well as the effective friction and diffusion.

In addition to this, it is possible to obtain a direct relation between the diffusion and drift.
An addition coupling term can be applied to the center of mass, which restricts it to a small region around a particular position $\vec{R}$ on the substrate.
The internal dynamics will not be affected, and only the average interaction [the first term in the second line of Eq.~(\ref{eq:driftPexpliciet})] term will change.
The effective friction and diffusion will therefore remain the same, and can be approximated as constants if the region is sufficiently small.
Hence, the center of mass is simply coupled to an infinite heat bath which produces diffusion and friction tensors ${\sf D}_{\mu\nu}$ and ${\sf G}_{\mu\nu}$ with
\begin{align}
{\sf D}_{\mu\nu}= \tilde{D}_{\mu\nu}(\vec{R}) + D_T \delta_{\mu\nu}~,\\
{\sf G}_{\mu\nu} = \tilde{\gamma}_{\mu\nu}(\vec{R}) + \gamma_T \delta_{\mu\nu}~.
\end{align}
The center of mass is coupled to an infinite heat bath consisting of the substrate and internal degrees of freedom of the molecule.
This heat bath necessarily has the same temperature as the substrate.
Consequently, similar to Eq.~(\ref{eq:Dgamma}) but in three dimensions, and allowing for the possibility of negative eigenvalues of the diffusion tensor,
\begin{align}
\tilde{\gamma}_{\mu\nu}(\vec{R}) = - \frac{1}{k_\mathrm{B}T} \sum_\kappa |{\mathcal D}_{\kappa}(\vec{R})| \, \vec{d}_{\kappa\mu}(\vec{R})\, \vec{d}_{\kappa\nu}(\vec{R})~,
\label{eq:Dgamma3D}
\end{align}
where ${\mathcal D}_\kappa(\vec{R})$ is the $\kappa$th eigenvalue of $\tilde{D}_{\mu\nu}(\vec{R})$, and $\vec{d}_{\kappa\iota}(\vec{R})$ is the $\iota$th component of the corresponding eigenvector. 
Equation~(\ref{eq:Dgamma3D}) is a particularly useful result, as Eqs.~(\ref{eq:difffinite}) and~(\ref{eq:fricfinite}) contain different information regarding the symmetries of the diffusion and friction tensors.
These symmetries can be used to improve the numerical evaluation of the theoretical expressions, and are used for this purpose in Sec.~\ref{sec:comparison}.

\subsection{Transport on the substrate}
The diffusion and friction tensors cannot, as yet, be measured directly in an experiment.
What we are interested in, therefore, is the combined average effects of all contributions to friction and diffusion, i.e. decay of correlation, on a typical trajectory on the substrate.
Besides the loss of correlation due to the thermal noise and the finite heat bath of the internal d.o.f.\ of the molecule, correlation can also decay due to chaotic dynamics in the slow coordinates.
As both diffusion coefficients and friction coefficients are additive, and the substrate is symmetric under rotations of 60$^\circ$, this can be written as
\begin{align}
\label{eq:sumgamma}
\gamma_{\mathrm{total},\mu\nu} & = \delta_{\mu\nu}(\gamma_{\mathrm{slow}} +\gamma_\mathrm{benzene} +\gamma_T)~,\\
\label{eq:sumD}
D_{\mathrm{total},\mu\nu} & = \delta_{\mu\nu} (D_\mathrm{slow} + D_\mathrm{benzene} + D_T)~.
\end{align}
where $\mu,\nu>3$, and the subscript slow and benzene indicate the contributions from the slow coordinates and internal d.o.f.\ of the molecule respectively.
Note that if the substrate lattice were rectangular, instead of hexagonal, the average friction and diffusion would not be isotropic.

As the invariant density is not affected by the noise level, and the fast dynamics are not affected strongly for up to realistic noise levels (1/ps), $\gamma_\mathrm{benzene}$ and $D_\mathrm{benzene}$ are constant for fixed temperature, irrespective of $\gamma_T$.
The slow dynamics are affected directly by the thermal noise, but this effect is small up to realistic noise levels.
They are also affected indirectly by the presence of the thermal noise, but not its intensity, through the invariant density, as can be seen from the first term in the second line of the right-hand side of Eq.~(\ref{eq:driftPexpliciet}), which contains the ensemble average.
The friction and diffusion of the reduced slow system, $\gamma_\mathrm{slow}$ and $D_\mathrm{slow}$ are not straightforward to treat analytically, even though this system is essentially a single particle moving in a complicated substrate potential (see, for instance, Ref.~\cite{vega2002}).

\section{\label{sec:md}Molecular dynamics simulations}

Molecular dynamics simulations were performed of the model system, and total friction and diffusion constants were calculated numerically.
In the simulated system, specific internal d.o.f.\ can be frozen, and the thermal noise due to the substrate temperature can be introduced slowly, thus providing insight into the effects of the internal d.o.f.\ and substrate temperature on transport.

The equations of motion based on Eq.(\ref{eq:hamiltonian}), with Eqs.~(\ref{eq:Vbenzene}) and~(\ref{eq:Vsubstrate}) substituted, were solved using the velocity-Verlet algorithm or in the case of Langevin damping and noise with a fourth order Runge-Kutta algorithm.
As this work is concerned with the dynamical properties of the system, it is necessary in the Hamiltonian case to use an algorithm which preserves the phase space volume exactly.
The largest Lyapunov exponent was obtained by numerically calculating the growth rate of an infinitesimal perturbation.
By virtue of the Oseledec theorem, this growth rate is equal to the largest Lyapunov exponent (see, for instance, Chap.~3 of Ref.~\cite{drieitalianen}).

Random initial conditions for the center of mass and the internal d.o.f.\ were chosen with energies corresponding to room temperature.
This was done by starting from a random point on the constant energy shell of the linearized system, with energy corresponding to room temperature and zero angular momentum, and subjecting the system to thermal noise for a period of time sufficient for reaching thermal equilibrium. 

The total friction coefficient 
was extracted from the trajectories by fitting a linear dependence to the average change in velocity as a function of the momentum, $\langle \vec{P}(t+\Delta t)-\vec{P}(t) \rangle_{\mathrm{traj}
}$.
Note that in order to accurately determine the friction due to the internal d.o.f., $\Delta t$ must be sufficiently long for the correlation in the internal d.o.f.\ to have decayed.
Similarly, the diffusion of the center of mass of the molecule on the substrate can be determined from the mean square displacement $\langle [\vec{R}(t+\Delta t)-\vec{R}(t)]^2\rangle_\mathrm{traj}$ as a function of the length $\Delta t$ of the time interval, as was done in, for instance, Refs.~\cite{onsbball,sokolovpre2004}.
If the diffusion is normal, $\langle [\vec{R}(t+\Delta t)-\vec{R}(t)]^2\rangle_\mathrm{traj}$ is proportional to $\Delta t$ for long $\Delta t$, and the prefactor is proportional to the diffusion coefficient.
If the mean square displacement does not grow linearly with time, but with some other exponent $\alpha$,
\begin{align}
\langle [\vec{R}(t+\Delta t)-\vec{R}(t)]^2\rangle_\mathrm{traj} \propto t^\alpha~,
\label{eq:anomalousalpha}
\end{align}
the diffusion is called anomalous.
Anomalous diffusion can be subdiffusion ($\alpha<1$) or superdiffusion ($\alpha>1$).
Note that anomalous diffusion cannot be described by a linear Fokker-Planck equation such as Eq.~(\ref{eq:fokkerplanck}).

\subsection{Fixed internal d.o.f.}

Besides the full system, several derived systems were considered with different sets of frozen internal d.o.f.\ with progressively fewer remaining internal d.o.f.\ of the benzene molecule, namely (1) frozen torsion modes, (2) all internal vibrations frozen while leaving rotation, and (3) fully frozen internal d.o.f.\ with (a) the molecule at incommensurate orientation, (b) the molecule at commensurate orientation, and (c) thermal average over all orientations.
Commensurate orientation means that the benzene molecule has the same orientation as the hexagons of C atoms in the substrate.
The results which are described in the rest of this section are summarized in Table~\ref{tab:tabel}.

The lowest three fundamental vibrational frequencies of the benzene molecule correspond to torsion modes, in which the C atoms oscillate in the $z$ direction.
The other modes have higher frequencies and involve exclusively bending and stretching.
By removing the dynamics in the $z$ direction, the torsion modes can be eliminated while leaving the other vibrational modes intact [case (1)].
Removing the torsion modes greatly affects the chaos in the internal degrees of freedom, because their coupling is strongly nonlinear and dominates the chaotic dynamics.
It also decreases the time-scale of the internal d.o.f., thereby increasing the separation between the time scales and decreasing $\epsilon$.
In molecular dynamics simulations without temperature, with a fixed center of mass position, and without the torsion modes, it was found that the largest Lyapunov exponent is typically below 0.014/ps depending on initial conditions.
This is much smaller than the value for the full system with fixed center of mass, 0.5/ps.
Without torsion, the largest Lyapunov exponent is, for some initial conditions, even zero within numerical error.
The resulting trajectories are not appreciably chaotic.
Note that due to the small time steps used in the simulations, the error in the Lyapunov exponents for a particular set of initial conditions is less than 0.001/ps.  There is, however, variation due to the variation in the initial conditions.

\begingroup
\squeezetable
\begin{table}
\caption{\label{tab:tabel}
Several dynamical properties for the full system and various sets of frozen internal d.o.f., namely the system without torsion, without any vibration, and without any internal d.o.f., for a fully frozen molecule at commensurate and incommensurate orientations, and for a thermal average over the orientation.
Shown are the largest Lyapunov exponents $\lambda$ of the noiseless system with and without a substrate in units of 1/ps at energies comparable to room temperature, the exponent of the diffusion $\alpha$, and the induced friction $\gamma-\gamma_T$ when the Langevin dynamics are switched on.
}
\begin{ruledtabular}
\begin{tabular}{lllll}
& {no substrate} & \multicolumn{3}{c}{{substrate}} \\ \cline{3-5}
&  & \multicolumn{2}{c}{{no noise}} & {noise} \\ \cline{3-4}
{internal d.o.f.} & $\lambda$ & $\lambda$ & $\alpha$ & $\gamma-\gamma_T$\\
\hline
full system & 0.5 & 1.2 & 1 & 0.83 \\
no torsion (1) & $0\leq\lambda\leq 0.014$ & 1.2 & $\geq 1$ & 0.77 \\
no vibration (2) & 0 & 1.2 & $\geq 1$ & 0.77\\
frozen, incomm. (3a)& --  & $<0.6$ & $1<\alpha<2$ & 0.29\\
frozen, comm. (3b)& --  & $<0.6$ & $1<\alpha<2.$ & 0.99\\
frozen, average (3c)& --  & $<0.6$ & $1<\alpha<2$ & 0.74 \\
\end{tabular}
\end{ruledtabular}
\end{table}
\endgroup

The bending and stretching modes have higher frequencies and weaker nonlinear coupling.
These modes can be frozen as well, by making the molecule completely rigid, leaving only one internal d.o.f., the rotation around an axis parallel to the $z$~axis [case (2)].
With only one internal degree of freedom, for fixed center of mass coordinates the molecule cannot be chaotic.
Because the fast, strong coupling between the atoms has been removed, the remaining dynamical system no longer has a fast time scale.
The time scale of the rotation is comparable with that of the center of mass motion.
Note that in very large rigid nanoclusters, the rotation is much slower than the center-of-mass motion~\cite{onsgraphiteflakes,onssquareflakes}.

For fully frozen internal d.o.f., the molecule is rigid [case (3)].
Depending on the initial orientation, the interaction with the substrate is different.
Therefore, both commensurate and incommensurate initial orientations are considered,
as well as systems where the interaction is equal to the thermal average for a rigid molecule that can still rotate.
This is essentially equal to the second line of Eq.~(\ref{eq:driftPexpliciet}) with the expression for the ensemble average in noisy systems, Eq.~(\ref{eq:thermalaverage}) for a rigid molecule with only rotational degrees of freedom, and a good first approximation of the thermal average for a molecule with all internal degrees of freedom.

\subsection{Trajectories\label{sec:trajectories}}

\begin{figure}
\includegraphics[angle=270,width=8.6cm]{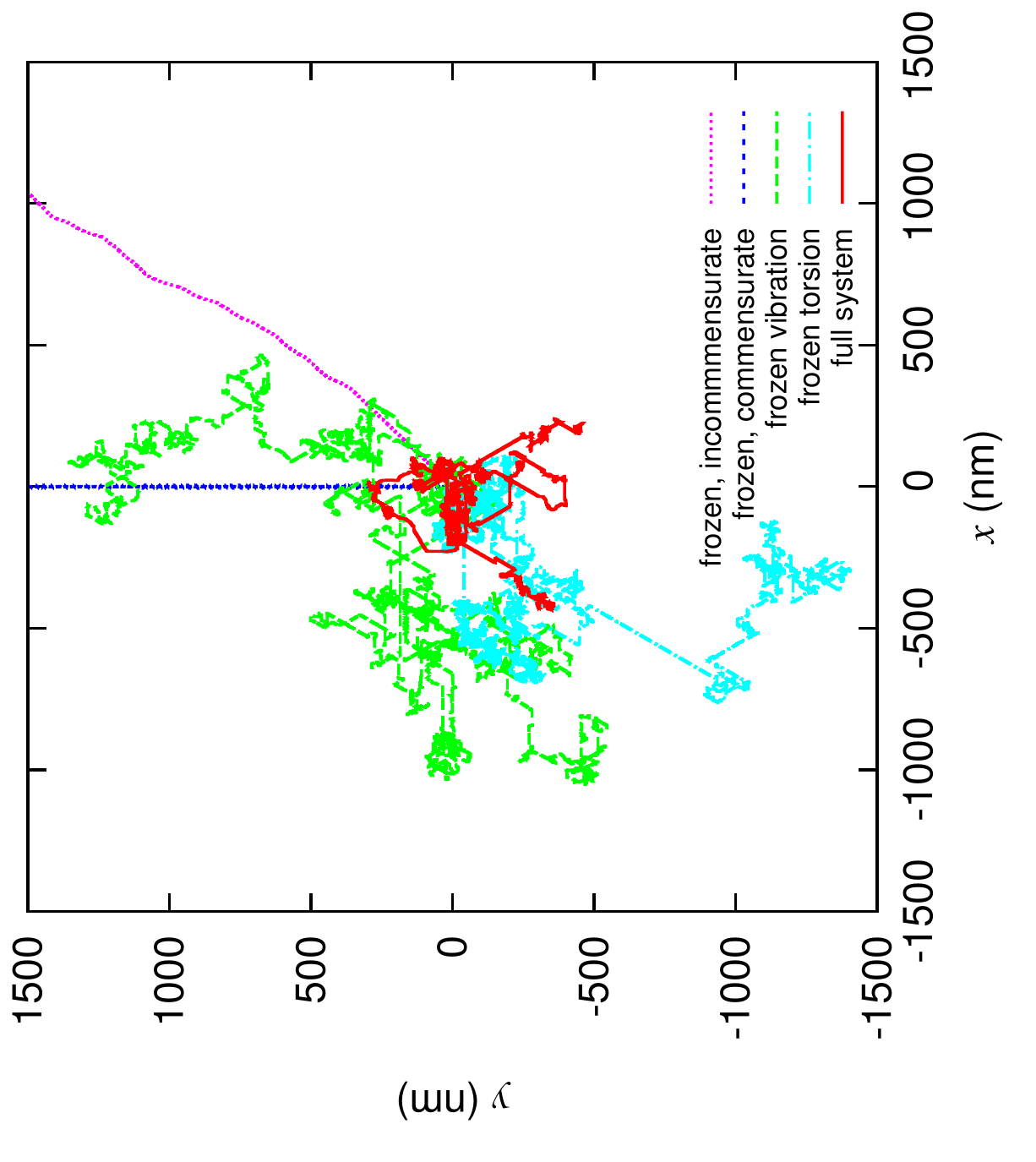}
\caption{
Simulated trajectories of the center of mass of a benzene molecule on graphite without temperature,
for the full system, as well as various systems in which some of the d.o.f.\ have been frozen.
In all cases the simulations lasted for 72~ns and initially the center of mass was at position $\vec{R}=(0,0,0)$.
Two distinct trajectories are shown for the full system, two trajectories for a system without motion along the $z$ direction, and two trajectories for a system with a rigid molecule which is allowed to rotate.
For fully frozen benzene molecules one trajectory is shown with incommensurate orientation and one with commensurate.
Long ballistic jumps can be seen even for the full system, which has normal diffusion.
\label{fig:trajectories}
}
\end{figure}

In Fig.~\ref{fig:trajectories}, examples of simulated trajectories of the center of mass on the substrate of equal duration are plotted for different dynamics of the internal d.o.f.\ and without thermal noise.
The full system is chaotic, with the largest Lyapunov exponent typically around 1.2/ps regardless of the precise initial conditions.
The motion of the center of mass on the substrate is diffusive, but is dominated by intermittent long-range ballistic motion.
The diffusion of the center of mass is normal with a total diffusion constant of around 27~nm$^2$/ns.

When the torsion modes are eliminated [case (1)], or when the internal d.o.f.\ are frozen completely except for rotation [case (2)], the dynamics remain chaotic.
The largest Lyapunov exponent in both cases is around 1.2/ps, similar to the full system.
The motion of the center of mass is diffusive.
For many initial conditions the diffusion is normal on the time scales of the simulations (120~ns), but for some initial conditions, superdiffusion cannot be excluded.
As the molecule itself in these two systems is either not chaotic, or very weakly chaotic, it is the addition of the substrate that leads to chaos.
In low-dimensional Hamiltonian systems with such a complicated phase-space structure, anomalous diffusion is to be expected for some initial conditions (see, for instance, Ref.~\cite{vega2002} and some of the chapters in Ref.~\cite{anomaloustransportbook}).

When all internal d.o.f.\ are removed [case (3)], the system becomes more weakly chaotic, with the largest Lyapunov exponent below 0.6/ps.
For some initial conditions, there is still diffusive behavior.
In nearly all of these cases, the diffusion is clearly superdiffusion, with scaling exponents $\alpha$ up to~1.9 [see Eq.~(\ref{eq:anomalousalpha})].

In the system under consideration here, the diffusion is not governed by single hops, but by ballistic motion in the directions that correspond to low diffusion barriers, as can be seen from Fig.~\ref{fig:trajectories}.  Ballistic motion was also observed in experiments~\cite{holly}.
The anomalous diffusion in the systems without torsion is also associated with these ballistic jumps.
In the full system, the noise from the chaotic dynamics of the torsion modes causes the ballistic trajectories to decay more quickly and reduces the total diffusion of the center of mass on the substrate to normal diffusion.
However, it is still dominated by long jumps, and, on short time scales, the diffusion is still anomalous, as has for instance also been found for single particles with Langevin dynamics in periodic and random potentials \hbox{\cite{sokolovprl2004,sokolovpre2004}}.

The results for the Lyapunov exponents and diffusion exponents are summarized in Table~\ref{tab:tabel}.

\subsection{Decay of correlation in the internal d.o.f.\label{sec:autocorrelation}}

\begin{figure}
\includegraphics[angle=270,width=8.6cm]{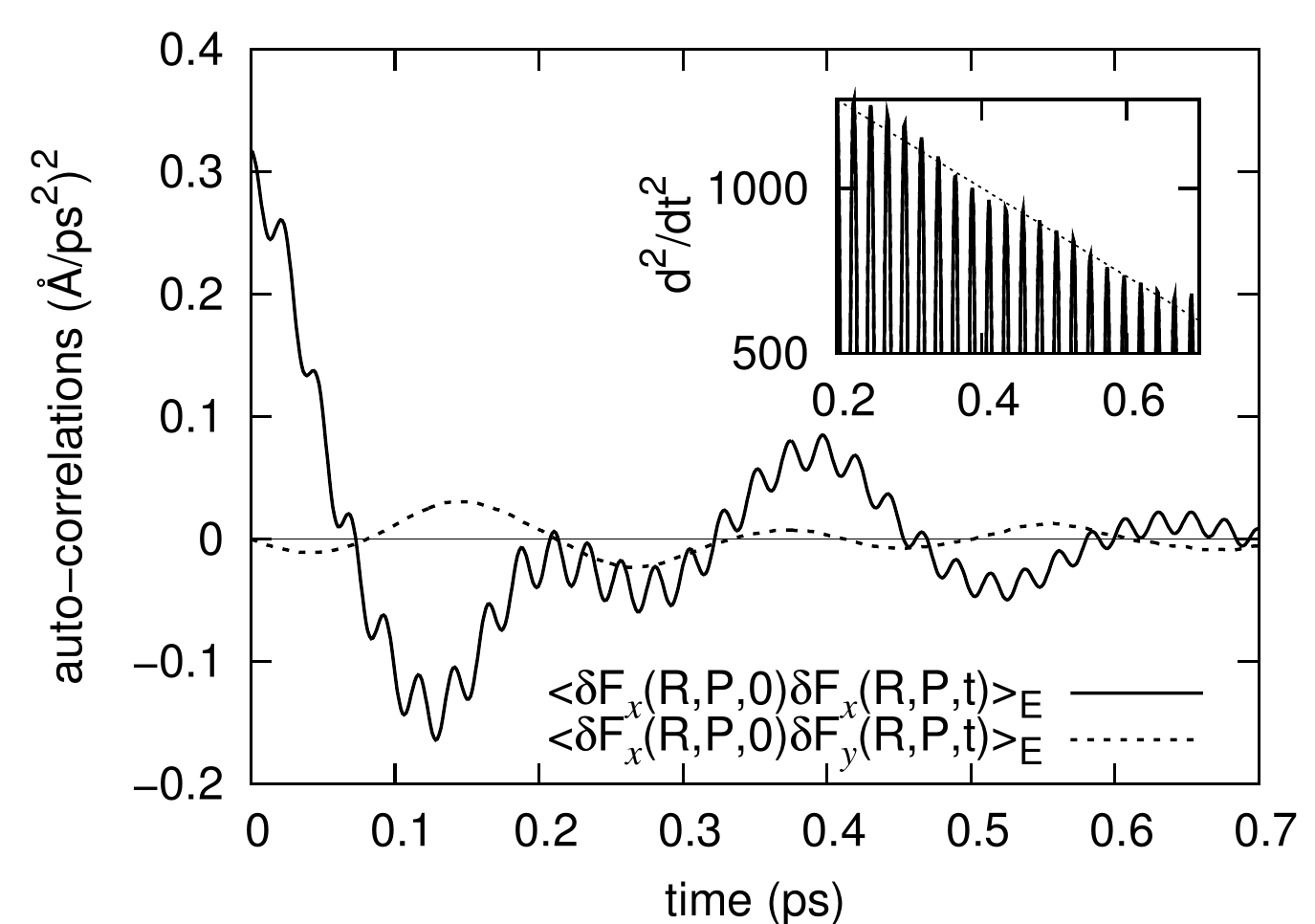}
\caption{
The auto-correlation function for various components of the coupling between the center of mass and internal d.o.f.\ as a function of time at $\vec{R}=(0,0,0), \vec{P}=(0,0,0)$ for a single simulated trajectory.
Due to the symmetries of the lattice in point (0,0,0), the auto-correlation functions for $y$--$y$ coincides with that for $x$--$x$.
From the theory, all elements must integrate to zero.
Modes with lower frequencies decay faster.
The inset shows a semi-log plot of the second derivative of the $x$--$x$ auto-correlation function, which emphasizes the high frequencies, and falls off exponentially.
The dotted line in the inset is an exponential fit to the maxima with a decay time of 0.55~ps.
\label{fig:autocorrelation}
}
\end{figure}

The effective drift and diffusion are affected by the correlation times of the internal d.o.f.,
as the effective momentum diffusion and friction due to the internal d.o.f., as expressed in Eqs.~(\ref{eq:difffinite}) and~(\ref{eq:fricfinite}), contain integrals over correlation functions of the coupling.
The auto-correlation function in the expression for the momentum diffusion, the integrand on the right-hand side of Eq.~(\ref{eq:difffinite}),
is straight-forward to evaluate numerically.
If 
the integral over the auto-correlation function is not finite,
the diffusion, if any, is anomalous and cannot be described by a linear Fokker-Planck equation such as Eq.~(\ref{eq:fokkerplanck}).

In Fig.~\ref{fig:autocorrelation} the auto-correlation function of the coupling is plotted for a benzene molecule kept at fixed position $\vec{R}=(0,0,0)$  and momentum $\vec{P}=(0,0,0)$.
It decays exponentially with a decay time of 0.55~ps.
The vibrations which persist for a long time correspond to the highest vibrational frequencies of the linearized system.
As the nonlinear coupling of these modes to other modes is weaker than that of the modes with lower frequency, correlations in them decay more slowly.
The time-integral of the autocorrelation function gives an estimate of the velocity-diffusion coefficient, via Eq.~(\ref{eq:diffusiontss}), which is equal to 0.0009~\AA$^2$/ps$^3$.
This value is close to 0, which is the value expected from Eq.~(\ref{eq:Dgamma3D}) in combination with the symmetries of Eq.~(\ref{eq:fricfinite}) in $\vec{R}=(0,0,0)$.
As the internal dynamics of the molecule are not strongly affected by the substrate, the decay in the internal d.o.f.\ is similarly exponential for all positions on the substrate.

\subsection{Thermal noise}

\begin{figure}
\includegraphics[angle=270,width=8.6cm]{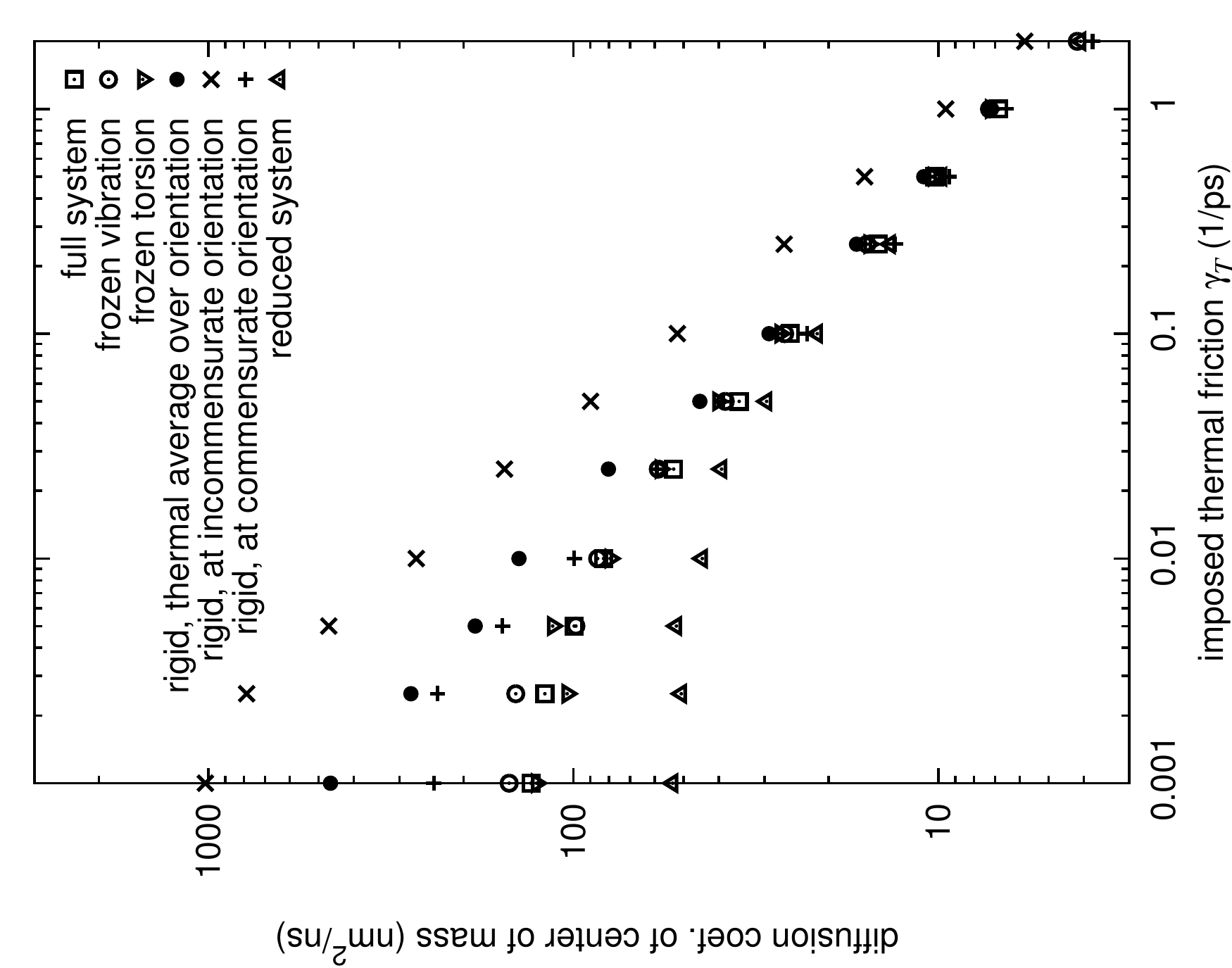}
\caption{
The diffusion coefficient of the center of mass on the substrate is plotted as a function of the thermal noise intensity (characterized by the imposed friction $\gamma_T$) in molecular-dynamics with Langevin dynamics at a temperature of 293~K.
Without Langevin dynamics, the systems which are totally frozen do not exhibit normal diffusion.
The error in the effective diffusion in the reduced system at low noise intensities is large, around factor of~3.
The diffusion coefficient of the full system without thermal noise is equal to 27~nm$^2$/ns.
The diffusion coefficients were obtained from trajectories with a total duration of 2~$\mu$s.
\label{fig:diffusion}
}
\end{figure}

We introduce thermal noise and vary the intensity between zero to realistic values, while setting the temperature equal to room temperature, as described in Sec.~\ref{sec:model}.
In Fig.~\ref{fig:diffusion}, the diffusion coefficient of a benzene molecule with various types of internal d.o.f.\ on the graphite substrate at room temperature is shown as a function of the imposed thermal friction $\gamma_T$.
As soon as the thermal noise is switched on, correlations in the fast subsystem decay exponentially, and diffusion becomes normal, regardless of the dynamics of the internal degrees of freedom.
Long ballistic jumps are destroyed faster by higher noise intensities, leading to slower diffusion.
At the realistic intensities corresponding to $\gamma_T=1$/ps, the diffusion of the center of mass for all systems which have the rotational degree of freedom is around 6.9~nm$^2$/ns, which, given the non-empirical nature of the substrate potential, is close to the experimental result of $5.39\pm0.13$~nm$^2$/ns~\cite{holly}.

With noise, the full system, the system without torsion, and the system without any vibration have very similar diffusion constants.
In these systems, the diffusion is slower than in systems without any internal d.o.f., due to the faster destruction of ballistic jumps.
In systems in which the diffusion is not dominated by ballistic motion, but by single hops between minima in the substrate potential, the additional noise due to the chaos in the internal d.o.f.\ might have more non-trivial effects, potentially enhancing the diffusion of the center of mass compared to thermal noise alone.

Of the three systems without internal d.o.f., the system with the molecule at incommensurate orientation has the highest diffusion coefficient, because it has the weakest interaction with the substrate.
At $\gamma_T \approx$~0.25/ps, the diffusion in systems with internal d.o.f.\ and systems without internal d.o.f.\ and average interaction has become nearly the same.
At this point, the noise due to the chaotic dynamics becomes weak compared to the thermal noise, and no longer affects the diffusion constant significantly.
As the expected noise intensity in the physical system corresponds to around 1/ps, this means that the diffusion of the benzene molecule is not significantly affected by the noise due to the finite heat bath of the internal degrees of freedom.
However, in systems where the time-scale separation parameter $\epsilon$ is larger, the effective momentum diffusion will be larger [see Eq.~(\ref{eq:diffusiontss})], and may become significant.
Furthermore, in the experiments of Ref.~\cite{longjumps}, long jumps dominate the diffusion even at realistic noise intensities, and consequently noise from internal degrees of freedom can modify diffusion significantly.

\begin{figure}
\includegraphics[angle=270,width=8.6cm]{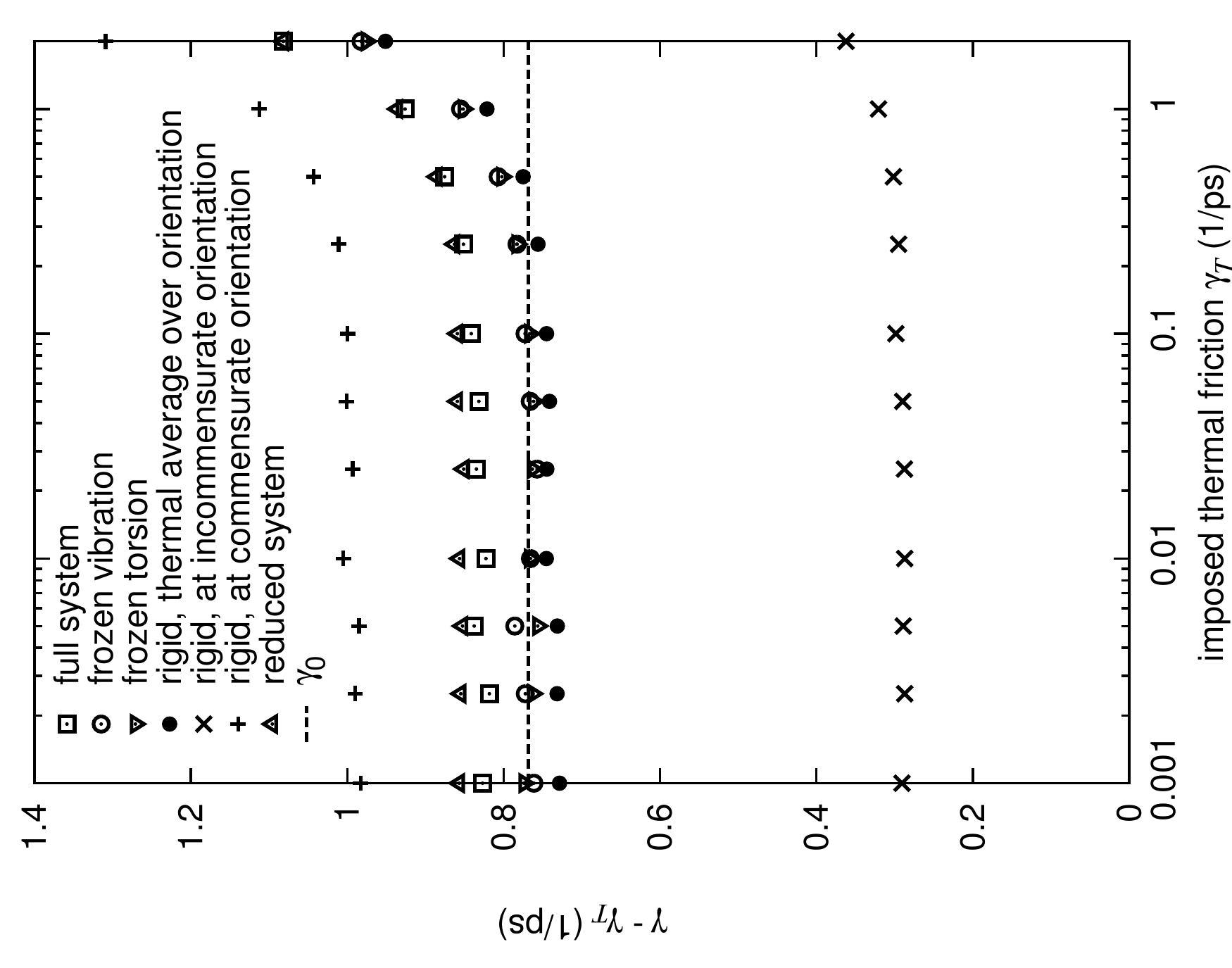}
\caption{
The component of the friction due to the interaction between the internal d.o.f.\ and the substrate is plotted as a function of the imposed friction in molecular-dynamics simulations with Langevin dynamics at a temperature of 293~K.
The friction coefficients were obtained by averaging over trajectories with a total duration of about $2~\mu\mathrm{s}$.
For comparison, the friction constant of the full system without thermal noise, $\gamma_0=0.75$/ps, is also plotted.
\label{fig:fric}
}
\end{figure}

In Fig.~\ref{fig:fric}, the induced friction coefficient is plotted as a function of the imposed friction.
The time interval used to estimate the friction was 0.484~ps, as most of the correlations in the internal d.o.f.\ have decayed after this time (see Fig.~\ref{fig:autocorrelation} and Sec.~\ref{sec:autocorrelation}).
At $\gamma_T=2.0$/ps, the total damping during the time interval of 0.484~ps is so strong that it becomes difficult to obtain an accurate estimate of the friction.  However, as can be seen from Fig.~\ref{fig:autocorrelation}, for shorter time intervals correlation has not decayed sufficiently to obtain accurate estimates.

\section{Comparison between theory, simulations, and experiment\label{sec:comparison}}

The qualitative theoretical results for the relation between internal chaos and diffusion compare well to the simulation results, as is summarized in Table~\ref{tab:tabel}.
In the full system, where the internal d.o.f.\ are chaotic (strictly positive Lyapunov exponent), and correlation decays exponentially, the noise from the internal d.o.f., combined with the time-scale separation, lead to normal diffusive behavior.
When the internal d.o.f.\ are not chaotic for some or all initial conditions, diffusion can be anomalous.
In the rest of this section, also quantitative results from the theory for the reduced system are compared to results from simulations of the full system.

\subsection{Evaluating the theoretical expressions numerically}
The theoretical expressions given in  Eqs.~(\ref{eq:difffinite}) and~(\ref{eq:fricfinite}) contain averages and integrals over correlation functions of the fast subsystem for fixed slow coordinates.
These expressions are difficult to evaluate in any more detail analytically, but here are evaluated numerically by performing simulations of the internal d.o.f.\ for fixed slow coordinates and calculating the necessary averages and correlation functions.
In these numerical simulations, the parameter $\gamma_T$ was set to a non-zero value, 0.1/ps, to ensure that the system is mixing and has the correct invariant density.

\begin{figure}
\includegraphics[width=8.6cm]{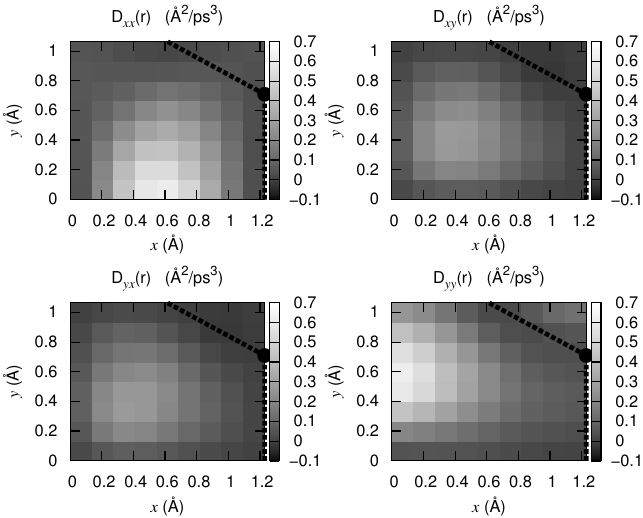}
\caption{
\label{fig:effdiff}
The effective velocity diffusion coefficient $\tilde{D}_{\mu\nu}(\vec{R})$ as a function of position, obtained by evaluating the theoretical expression, Eq.~(\ref{eq:difffinite}), numerically.
For reference, carbon atoms and bonds between them are indicated with solid circles and dotted lines.
From Eq.~(\ref{eq:difffinite}) it is clear that the effective momentum diffusion does not depend on the velocity or on the $z$ coordinate.
}
\end{figure}

The average coupling term [the first term of the second line of Eq.~(\ref{eq:driftPexpliciet})] is simple to evaluate in numerical simulations.
Similarly, the auto-correlation function needed for the momentum diffusion, shown in Eq.~(\ref{eq:difffinite}), is straight-forward to obtain from simulated trajectories and to integrate.
The results for the effective momentum diffusion as a function of the position are shown in Fig.~\ref{fig:effdiff}.

\begin{figure}
\includegraphics[width=8.6cm]{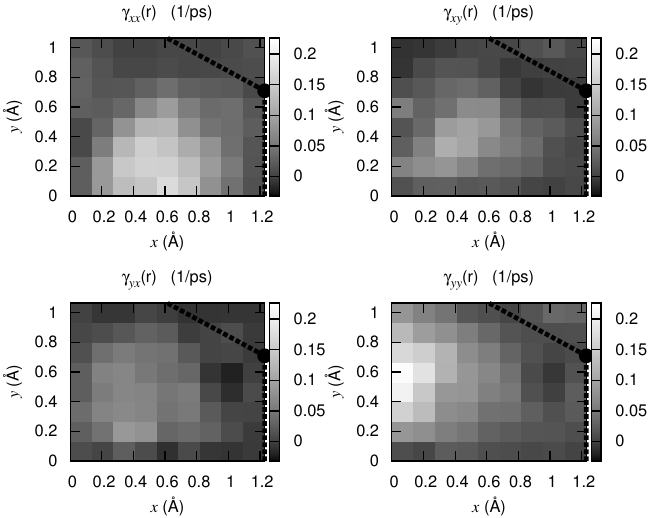}
\caption{
\label{fig:efffric}
The effective friction coefficient $\tilde{\gamma}_{\mu\nu}(\vec{R})$ as a function of position, obtained by evaluating the theoretical expression, Eq.~(\ref{eq:fricfinite}), numerically.
Autocorrelations were calculated for up to 0.7~ps and integrated. 
For reference, carbon atoms and bonds between them are indicated with solid circles and dotted lines.
As with the diffusion, the effective friction coefficient does not depend on the velocity or the $z$ coordinate [see Eq.~(\ref{eq:fricfinite})].
As it is more computationally involved to calculate the friction tensor than the diffusion tensor, the results are less accurate.
Within the error of the numerical evaluation, the effective friction indeed has the same shape as the diffusion, shown in Fig.~\ref{fig:effdiff}, with the correct ratio, and thus obeys Eq.~(\ref{eq:Dgamma3D}).
}
\end{figure}

The expression for the friction coefficient [Eq.~(\ref{eq:fricfinite})] is more complicated to evaluate quantitatively, due to the derivative.
Because the invariant measure depends on the position of the center of mass, the average and the derivative are not interchangeable. 
Hence, in order to calculate the derivative, the average coupling must be evaluated sufficiently accurately at nearby positions.
Numerical evaluation of this expression is therefore more time consuming than for the diffusion, and less accurate.
In order to obtain sufficient statistics, the nearby positions had to be chosen to be 0.26~\AA\ apart.
Results for the effective friction coefficient are shown in Fig.~\ref{fig:efffric}.

As can be seen from Figs.~\ref{fig:effdiff} and~\ref{fig:efffric}, within the error of the numerical evaluation, the effective momentum diffusion and friction indeed have the same shape, with the correct ratio, and thus obey Eq.~(\ref{eq:Dgamma3D}).

\subsection{Comparison of theory with simulations of the full system}
It is possible to compare the theoretical results to numerical trajectories.
However, comparing the results from Eqs.~(\ref{eq:difffinite}) and~(\ref{eq:fricfinite}) that were obtained numerically in the previous subsection to diffusion and drift tensors obtained directly from numerical simulations of the full system requires an accurate determination of the mean square displacement in velocity space from trajectories of the entire system.
However, due to the high dimensionality of the slow subsystem, extremely long trajectories are needed to obtain useful statistics.
With currently available computing power this is not yet possible.
It is interesting to note that the theory can thus be used to calculate properties of the system that cannot yet be obtained from molecular-dynamics simulations.

Nevertheless, the theory can be tested by comparing total diffusion and friction coefficients of the center of mass on the substrate found from numerical simulations of the reduced dynamics to those obtained from simulations of the full system.

The numerical values for the theoretical expressions at specific points, obtained in the previous subsection, can be used to obtain an approximation for the effective drift and diffusion, by fitting it to a Fourier expansion with the appropriate symmetries, cut off after a certain wave vector (in this case $8\pi/a$).
These symmetries are the translation, rotation, and reflection symmetries of the lattice, the reflection symmetries of Eq.~(\ref{eq:fricfinite}), the symmetry of the diffusion tensor in Eq.~(\ref{eq:difffinite}), and finally Eq.~(\ref{eq:Dgamma3D}).
Though, in principle, there are many terms, due to the large number of symmetries only four free parameters remain.

Results for the total friction and diffusion of the reduced system are included in Figs.~\ref{fig:diffusion} and~\ref{fig:fric}.
At low noise levels, the induced friction in the full system is about 0.83/ps, while the reduced system has an induced friction of $(0.86\pm0.02)$/ps.
The error
is due to the error in the (fit to the) numerical evaluations of the effective momentum diffusion and friction and was estimated with a variational method.

The diffusion of the center of mass in the reduced system follows that of the full system for high noise levels.
For low noise levels, however, there is a discrepancy.
This is due to the fact that in this regime, the diffusion is dominated by long ballistic jumps, as was mentioned in Subsec.~\ref{sec:trajectories}.
The long jumps are very sensitive to noise and damping, and disappear at high noise levels.
The noise levels needed to destroy the long jumps are comparable to the typical intensities of noise and damping from the internal degrees of freedom.
Nevertheless, the long jumps survive in both the full and reduced system, because for some positions on the substrate there are no noise and damping from the internal degrees of freedom.
The duration of the jumps, however, depends sensitively and in a highly nontrivial way on the reduced dynamics in these regions, where the noise level due to the internal d.o.f.\ is extremely low.
Due to the statistical errors in the numerical evaluation of the theoretical expressions, the error in the effective momentum diffusion at these noise intensities is large, around a factor of 3.
It should be noted that while the thermal damping $\gamma_T$ is usually taken to be a constant, it too could, in principle, depend on position, which could affect the survival of long jumps.

\subsection{Various internal degrees of freedom}

As the intensity of the thermal noise affects the dynamics of the internal d.o.f.\ only weakly, and not the invariant density, the effective friction and momentum diffusion are not affected by it.
From Fig.~\ref{fig:fric},
it can be seen that the induced friction, which contains the effective friction due to the finite heat bath of the internal d.o.f., does not depend on the noise intensity.
Together with the values for the induced friction in the various systems, one can use this to separate out the contributions to from the internal degrees of freedom.

The substantial difference between the induced frictions in the fully frozen systems with commensurate and incommensurate orientations (0.99/ps and 0.29/ps respectively) demonstrates that the interaction between the internal d.o.f.\ of the molecule and substrate strongly affects the friction.
The full system and other systems with reduced or partially frozen dynamics lie somewhere in between these two extremes.
The next best approximation, the system without any internal d.o.f., but with the coupling averaged with thermal weight over orientation has an induced friction coefficient of 0.74/ps.
Because it is not related to the internal d.o.f.,
this amount can be attributed to the slow dynamics.

The induced friction in the full system is 0.83/ps for a wide range of noise intensities, whereas, without thermal noise the full system has an induced friction of $\gamma_0=0.77$/ps.
This difference is due to the fact that even weak thermal noise changes the invariant density significantly compare to the system without noise (see Sec.~\ref{sec:noise}) and this affects the friction induced by the finite heat bath [see Eq.~(\ref{eq:drifttss})].

The induced friction in the systems which still possess rotation, but have frozen torsion modes, is 0.77/ps in both the system with and without bending and stretching.
The difference between this value and the value for the full system can be attributed to the chaotic dynamics of the torsion modes.
The remaining difference to the system with a rigid molecule with thermal average over orientation is due to rotational dynamics.

In summary, three contributions to the total friction can be identified: the damping due to the thermal heat bath of the substrate ($\gamma_T\approx 1$/ps), a constant contribution from the slow dynamics ($\gamma_\mathrm{slow}=0.74$/ps), and the contribution from the chaotic internal d.o.f.\ ($\gamma_\mathrm{benzene}=0.09$/ps), which is dominated by the torsion modes.
This contribution from the internal d.o.f.\ is a combination of the average coupling and finite heat bath due to the internal vibrations.

The total induced friction of 0.83/ps found in simulations matches well with the theoretical predictions from the time-scale separation theory, $0.86\pm 0.02$/ps.
With the addition of the 1/ps substrate friction due to dissipation into phonon modes, the numerical and theoretical calculations compare favorably with the result of $2.2\pm 0.1$/ps from He/neutron spin echo experiments~\cite{holly}.

\section{\label{sec:discussion}Conclusions}
In this work, the relationship between the internal d.o.f.\ of a benzene molecule and its motion on a graphite substrate were investigated using time-scale separation theory and dynamical properties, including the effects of thermal fluctuations.
It was shown that the fast dynamics of the internal d.o.f.\ act as a finite heat bath, while the motion of the center of mass is slow.
It was found that thermal noise affects the internal dynamics of the molecule weakly,
yet ensures the requirements of the time-scale separation theory of Ref.~\cite{tss1}, i.e.\ exponential decay of correlation and mixing.
The invariant density of the internal d.o.f.\ in the presence of noise is equal to the thermal distribution.
The total friction and  diffusion of the center of mass consist of contributions from the thermal heat bath of the substrate, the finite heat bath of the molecule, and the decay of correlation in the slow dynamics.

The theoretical expressions for the effective dynamics were evaluated numerically, and numerical simulations of the reduced system were performed.  Their results compare well with those from simulations of the full system.
In the molecular dynamics simulations, several sets of internal d.o.f.\ were frozen.
It was found that the torsion of the benzene molecule dominates the chaotic dynamics.
Without them, there is either anomalous superdiffusion, or ballistic motion.

The total friction at room temperature consists of the damping due to the infinite heat bath of the substrate, a contribution from the slow dynamics, and one from the internal degrees of freedom.
These contributions together are sufficiently large to account for the high friction found in experiments~\cite{holly}.

The present results suggest several approaches which may make it possible to utilize the internal d.o.f.\ to affect diffusion and friction on a substrate.
Molecules which are floppy possess many low-frequency vibrational modes with strong nonlinear coupling, such as the torsion modes in benzene, and consequently, their internal d.o.f.\ produce more noise.
Furthermore, as the effective drift and diffusion decrease with the time-scale separation parameter, combinations of molecules and substrates where the time-scale separation is weaker will produce larger effects.
Finally, by selectively exciting vibrational frequencies of the adsorbate with laser light or by other means, the total energy in the internal d.o.f.\ could be increased, enhancing the noise due to the finite heat bath, and therefore changing the effective momentum diffusion and friction.
In this context, large flat photo-active molecules such as phthalocyanines are particularly interesting.

\begin{acknowledgments}

The author is grateful to A. Fasolino, S. Hallerberg, and A. Rowan for stimulating discussions.
The work has been financially supported by a Veni grant of Netherlands Organization for Scientific Research (NWO).

\end{acknowledgments}

\end{document}